# A modeling system for monitoring the air - water pollution

V. F. KRAPIVIN[1], C. A. VAROTSOS[2] and B. Q. NGHIA[3]
*Kotelnikov Institute of Radioengineering and Electronics, Russian Academy of Sciences, Moscow, Russia[1] .*
*Department of Environmental Physics and Meteorology, University of Athens, Athens, Greece[2].*
*Vietnam Institute of Logistics, Ho Chi Minh City, Vietnam[3] .*

**Abstract.** Regular monitoring of key water quality parameters is important for assessing its the hydrological status in conjunction with air-pollution interaction. In this study, a new cost - effective technique based on the geo-ecological information-modeling system (GIMS) is implemented employing the combined use of simulation experiments and in field observations to investigate the problem of optimizing water quality monitoring. The GIMS, is accompanied by 39 elements selected in 8 management systems and 31 functional elements, which are described in detail. It is shown that the combined use of model and field observations allows reliable recording of water quality and optimization of the monitoring regime. Finally, simulation experiments in a lagoon are presented, demonstrating the fidelity of the proposed modelling system to optimize water quality control through regular in field measurements and simulations.

## 1. Introduction

There is no doubt that human activities sometimes have devastating effects on the environment [1-3]. The composition of the atmosphere changes over time not only in quality but also in the quantity of its components. The causes of these changes are both natural and anthropogenic [4-6]. The percentage contribution of each component is not always known with sufficient accuracy. A classic example the last four decades is the ozone depletion and its feedbacs [7-13]. Ground-based, satellite-borne, balloon and aircraft instrumentation gathered a large dataset to study the global ozone concentration variability [14, 15]. Undoubtedly, the anthropogenic contribution to the decrease of the ozone amount in the stratosphere and the increase of ozone in the troposphere has proven experimentally [16-21]. The latter contributes greatly to the photochemical smog effect of the megacties and the temperature structure in the troposphere [22-27].

Human activities, apart from the deterioration of the atmosphere, also alter the quality of water. For example, according to current estimates, millions of tons of manure are produced annually, which is often stored for long periods in tanks or lagoons. This activity causes additional atmospheric pollution with various hazardous gases, like H2S, CH4, NH3 and CO2, which among others enhance the atmospheric greenhouse effect (e.g. [28-30]).

Nuoc Ngot Lagoon is located in South Vietnam (Binh Dinh Province, $14°9'0''$N, $109°10'59''$E) in the zone where there are intense anthropogenic activities, which are mainly related to agriculture [31-33]. The importance of the lagoon ecosystem is determined by its productivity that is assessed by fish and shrimp fishery. Protection of the high efficiency of the lagoon ecosystem takes place through a traditional monitoring system that includes in-situ measurements of the water quality. A geo-ecological information-modeling system (GIMS) that includes GIS-technology functions has been recently proposed [34]. This paper adopts the GIMS and applies it in the Nuoc Ngot Lagoon area.

The basic function of GIMS is the optimization of the monitoring process in order to integrate and coordinate the collection of monitoring data obtained from various instrumentation tools in episodic time periods and certain regions in space.

The structure of GIMS depends on the complexity of the investigated natural field. The Nuoc Ngot lagoon area is characterized by a variety of the soil-plant configurations, including their ecological features and the functions of the climate parameters. The regime of lagoon water is controlled by the exchange between the systems of South-China Sea and the river.

The GIMS adaptation in the lagoon area requires the development of new algorithms along with the specification of the standard GIMS items, such as instrumental measuring tools and a regional hydrology model.

There are numerous studies discussing the heat budgets, the energy, and the hydrological regimes of coastal lagoons. In any case, these studies compose specific models to provide the basis for optimizing the monitoring data and the processing techniques. In this context, a natural heat budget model for the Nueva Lagoon in southeast Spain has been developed [35]. This model is only connected with the monitoring process which is suitable for the Nueva Lagon area. In addition, the development of a model for the Lesina Lagoon (Italy} based on artificial neural networks to study the balance of lagoon water and to assess the spatial variability of lagoon biogeochemistry has been explored [36]. By using traditional modeling technique the Curonian Lagoon hydrological regime as a function of the modernization projects of the Klaipėda Seaport was analysed [37], taking into account climate change and the influence of the Baltic Sea. These and other similar models are characterized by the individuality of the model structure and its specific functions.

Direct use of the above-said models for studying other lagoons requires additional investigations and the development of a new model. The GIMS simplifies the solution to this task. The structure and functions of GIMS are oriented to the variety of parameters of a typical environmental system that allow the specification of the components of the nature-society system. These are typical for given spatial and temporal scales which characterize the environmental system under investigation. In this case, a stage of the model development comes into the GIMS adaptation process, when the model is automatically synthesized through the formation of subject identifiers in the GIMS database. In particular, the GIMS structure consists of five levels of functional elements, which can briefly described as follows:. Elements of the first level represent the regional image of the studied environmental system. The second-level elements is structured with processing algorithms of monitoring data. Third-level elements form a spatial image of the studied environmental system that lists the system elements in spatial digitization pixels. Fourth level items produce links between GIS structures and models. Five level items provide the forecasting process and an assessment of the simulation results.

## 2. Model description and its implementation: The GIMS structure and functions adopted

According to the general technique employed for the synthesis of GIMS, it is necessary to create a simulation model of the studied environmental object. It should be done taking into account the *a priori* relevant information and the assessment of the deviations between the model's results and the observations.

This allows the correction of the model and monitoring regime. As a result, a consistent coordination of the model calculations and monitoring data is established and a reliable forecast is made. The standard GIMS structure has a set of models shown in Figure 1 and Tables 1 and 2.

The GIMS adaptation process is performed by the GSI item which configures the hydrospherical part of the lagoon area under investigation and produces a universal schematic envelope of the lagoon zone covering the area with a grid $\Delta\varphi$ in latitude and $\Delta\lambda$ in longitude provided by the user. A discrete numbering of pixels $\Xi_{ij}=\{\varphi_i\leq\varphi<\varphi_{i+1}, \lambda_j\leq\lambda<\lambda_{j+1}\}$ gives a set of locations, where each location $\Xi_{ij}$ has an area $\sigma_{ij}$. The introduced grid of digitizing the lagoon zone is identified by a series of subject identifiers that correspond to the lagoon elements. According to this procedure, the lagoon zone $\Xi$ in the GIMS/NNL database is presented by a set of matrices $A_k=\left\|a_{ij}^k\right\|$, where $a_{ij}^k$ is the identifier of the subject or process in the pixel $\Xi_{ij}$.

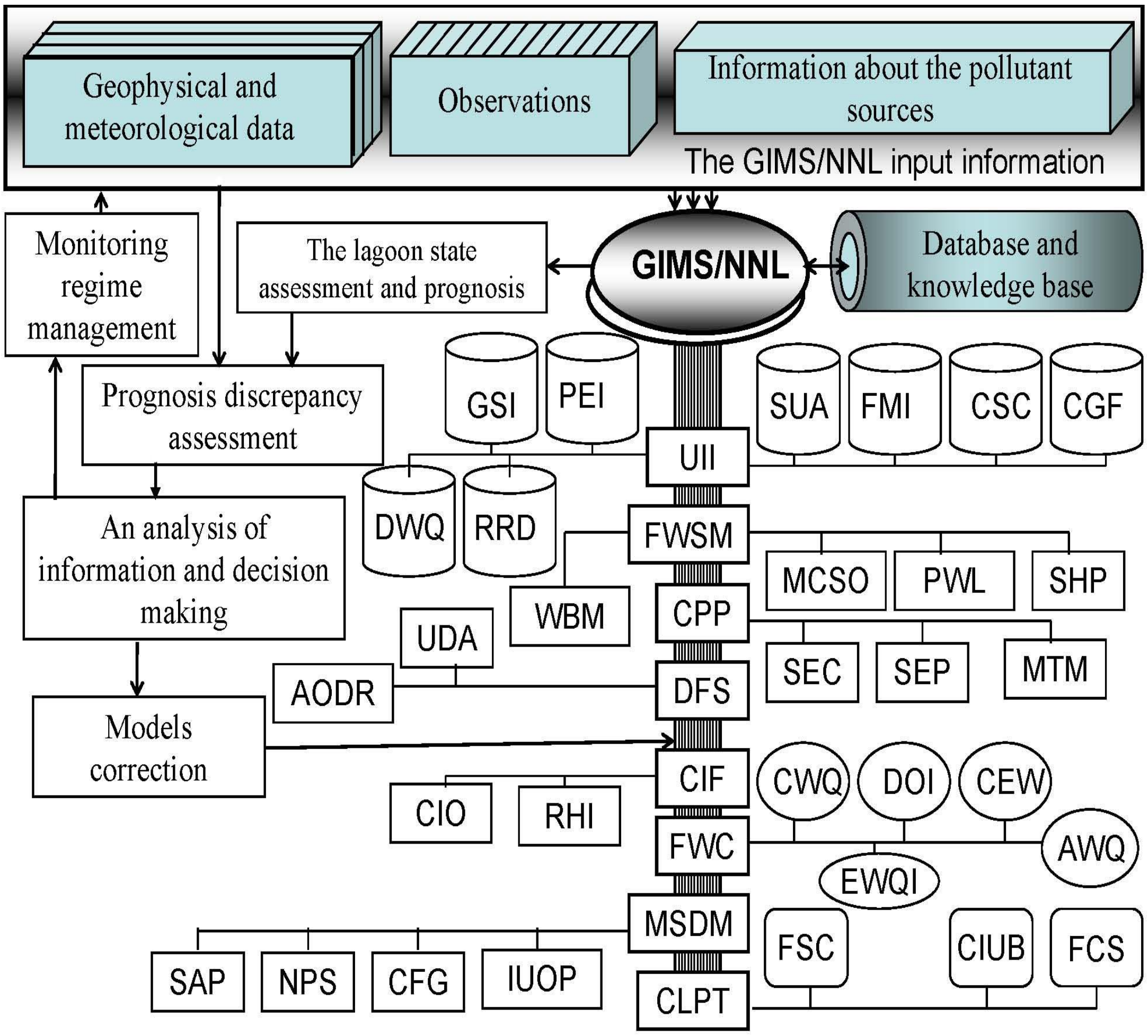

**Fig. 1.** Structure of the geo-ecological information-modeling system of the Nuoc Ngot Lagoon (GIMS/NNL). The system items are described in Tables 1 and 2.

**Table 1**. The GIMS/NNL management items.

| Item | Item functions |
|---|---|
| UII | Universal information interface. |
| FWSM | Formation of the water balance simulation model. Management of the models and algorithms for the parameterized description of the hydrophysical, hydrochemical and hydrological processes. |
| CPP | Control of the parameterization process for the energy and matter flows in the lagoon. Realization of the transformation mechanisms for chemicals in the water environment. |
| DFS | Database formation and synthesis of anthropogenic scenarios which are likely in the lagoon area. |
| CIF | Control of information flows between the GIMS/NNL items. |
| FMI | Formation and use of the water quality criteria.. |
| MSDM | Management of the statistical decision makings. |
| CLPT | Control of the lagoon phase transitions. |

In this regard Fig 2 shows the main stages of the lagoon zone identification and the water quality assessment.

The hydrological regime of the lagoon is main component of environmental processes that determine the water quality. The items WBM and MCBO describe the typical hydrological processes in the lagoon zone, taking into account the interaction between the lagoon and the South-China Sea. This interaction is associated with water exchange at the lagoon mouth Γ (strait). Figure 3 represents a set of water flows that are possible in each pixel of the lagoon zone. The water cycle is the continuous circulation of water in the lagoon zone, described by the system of differential equations taking into account the external water flow and a drift water flow.

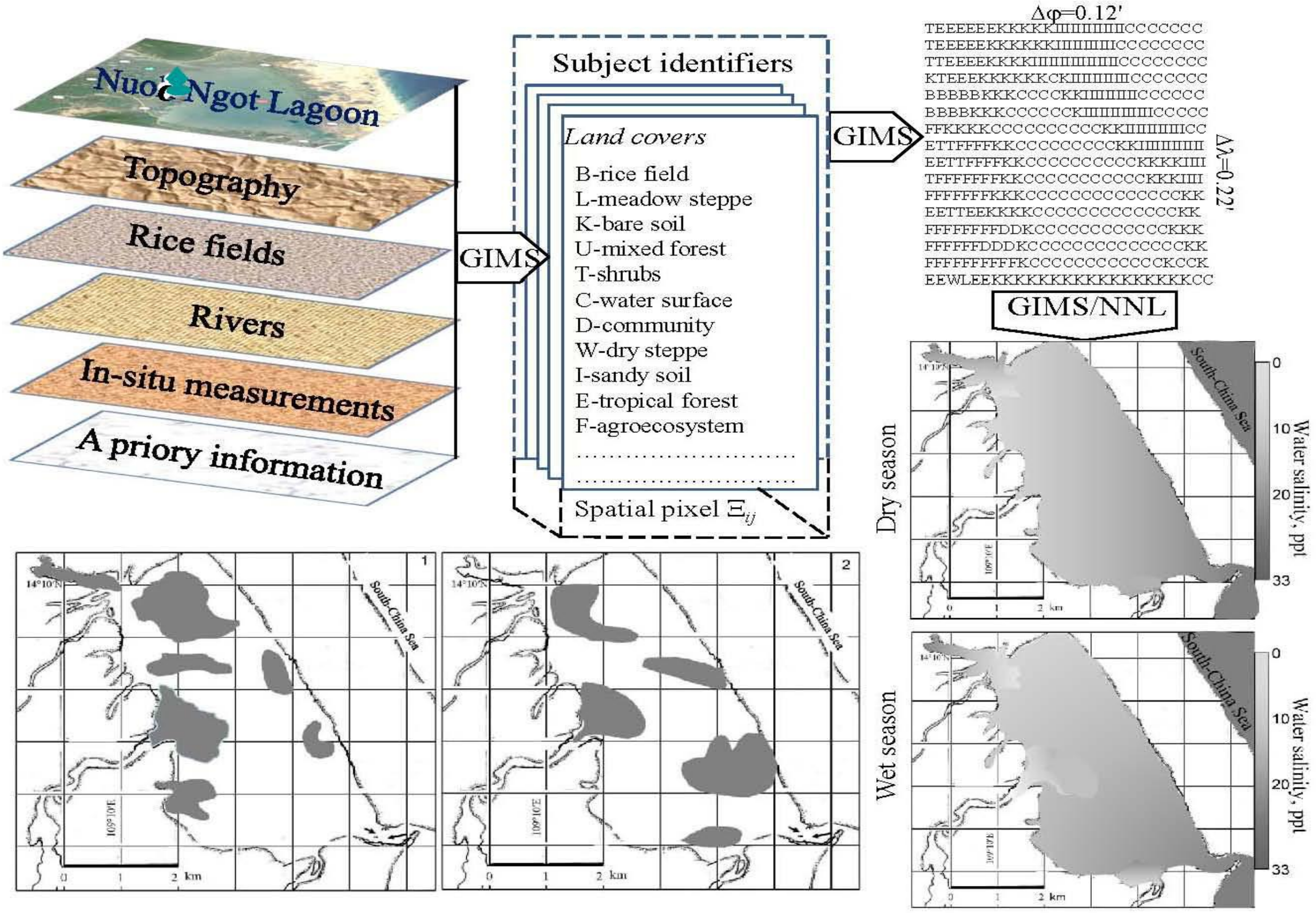

**Fig. 2.** The GIMS/NNL application for the Nuoc Ngot Lagoon monitoring and water quality diagnostics. Note: (1), $NO_3^-$ concentration exceeds 0.25 ppm; and (2), pH ≥ 9.0.

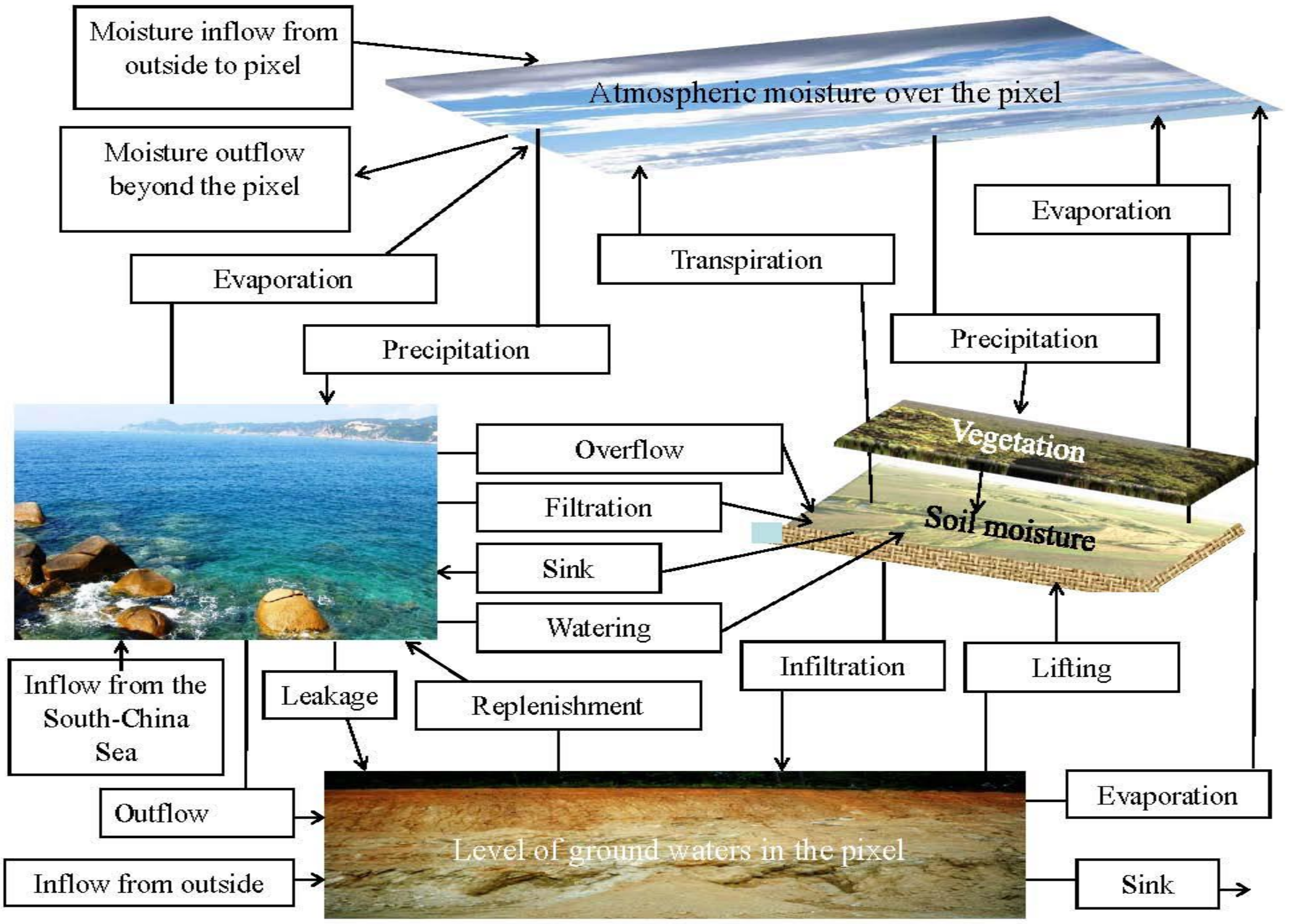

**Fig. 3.** The block diagram of the model for the water balance in each pixel of the lagoon zone.

**Table 2.** The GIMS/NNL functional items.

| Item | Item functions |
|---|---|
| GSI | Generation of the subject identifiers to adopt the GIMS to the lagoon region configuration taking into account of geophysical, ecological and socio-economic structure. |
| PEI | Perception of experimental information, its scaling and the entry to the database. |
| RRD | Realization of the requests to the database. |
| SUA | Supporting the user actions when a decision about the interface corrections are needed.. |
| FMI | Forming the maps with information about water quality in the lagoon. |
| CSC | Change of the scales for the cartographic information with the selection of the lagoon bordering territory. |
| CGF | Control of the GIMS/NNL functions to provide the coordination of interial information fluxes, to detect the defective requests and messages, to notify about the incorrect or unlawful user commands, to support the user actions |
| DWQ | Detection of the water quality disturbances and the user informing about it. |
| WBM | Water balance model of the lagoon influence zone [34]. |
| MCSO | Model of complex multi-factor surface outflow taking into account of the catchment area topography and soil-plant formations. |
| PWL | Parameterization of the wastewaters to the lagoon [38, 39] |
| SHP | Simulation of hydrophysical processes [34, 40]. |
| EWQI | Evaluation of the water quality indicators [41-43]. |
| MTM | Modelling the transformation mechanisms of the chemicals in the water environment [40]. |
| SEP | Simulation of the exchange processes on the boundary lagoon-sea including the tide [35-44]. |
| SEC | Simulation of the exchange processes by chemicals between the lagoon and atmosphere [45]. |
| UDA | Updated data archive of pollutant characteristics that can be delivered to the lagoon zone from the agriculture, municipal and industrial sources. |
| AODR | An assessment of official data reliability concerning the sources of pollutants. |
| RHI | Reduction of the heterogeneous information to the unique standard. |
| CIO | Coordination of the inputs and outputs of the GIMS/NNL items and their connections with the database. |
| CWQ | Control of the water quality criteria. |
| DOI | Documentation of operative information concerning the lagoon water quality. |
| AWQ | Accounting the water quality analyses realized in the chemical laboratory. |
| CEW | Complex evaluation of the lagoon water quality. |
| NPS | Neyman-Pearson statistical decision making procedure.. |
| SAP | Sequential analysis procedure to make the statistical decisions. |
| CFG | Control of the functioning the GIMS/NNL items. |
| IUOP | Information uncertainty overcoming procedure. [46]. |
| FSC | Formation of the series for meteorological and geophysical characteristics that are specific for the lagoon zone. |
| CIUB | Calculation of the indicators that characterize an instability of the environment and the lagoon ecosystem biocomplexity. |
| FCS | Forming the cluster space of the lagoon water quality features. |

The process of diffusion of contaminants into lagoon water depends on their condition and takes place with the process of water circulation. Dissolved fractions of contaminants ($\xi$) are more involved in biogeochemical processes than in suspended particles ($\mu$) [46]. But as suspended particles, the contaminants fall faster into the sediment. Therefore, the MTM item describes the processes of transformations of contaminants, including absorption of the dissolved fraction $\xi$ from the plankton ($H_{Z\xi}$), the sedimentation of the solid fraction ($K_{1\mu}$), the deposition with the detritus ($H_{D\xi}$), the absorption by detritophages from the bottom sediments ($K_L^{\mu\xi}$), and release from the bottom sediments owing to the diffusion ($K_a^{\mu\xi}$). As a result, the dynamic equations describing the contaminant cycle in the lagoon environment become:

$$\frac{\partial \mu_w}{\partial t} + v_\varphi^w \frac{\partial \mu_w}{\partial \varphi} + v_\lambda^w \frac{\partial \mu_w}{\partial \lambda} + v_z^w \frac{\partial \mu_w}{\partial z} = \sum_{i=1}^{3} \beta_i \Omega_{\xi\mu}^i - K_{1\mu} + \alpha_1 K_a^{\mu\xi}$$

$$\frac{\partial \xi_w}{\partial t} + v_\varphi^w \frac{\partial \xi_w}{\partial \varphi} + v_\lambda^w \frac{\partial \xi_w}{\partial \lambda} + v_z^w \frac{\partial \xi_w}{\partial z} = (1-\alpha_1) K_a^{\mu\xi} + k_2^w \frac{\partial \xi_w}{\partial z^2} - H_{Z\xi} - H_{D\xi} - H_{a\xi}$$

$$\frac{\partial \mu^*}{\partial t} = K_{1\mu} - \alpha_1\left(K_L^{\mu\xi} + K_a^{\mu\xi}\right), \qquad \frac{\partial \xi^*}{\partial t} = H_{D\xi} - (1-\alpha_1)\left(K_L^{\mu\xi} + K_a^{\mu\xi}\right)$$

where $\mu^*(\mu_w)$ and $\xi^*(\xi_w)$ are the concentrations of contaminants in the bottom sediments (water) as solid and dissolved phases, respectively; $H_{a\xi}$ is the output of contaminants from the sea to the atmosphere by evaporation and spray; $\Omega^i_{\xi\mu}$ is the input of contaminants into the lagoon with river waters ($i$=1), atmospheric deposition ($i$=2) and ship's wastes ($i$=3); $\beta_i$ is the part of the suspended particles in the $i$th flow of contaminants; $V = \{v^w_\varphi, v^w_\lambda, v^w_z\}$ is the flow velocity, and $\alpha_1$ is the part of the solid fraction of contaminants in the bottom sediments.

As expected, salinity of water is an important feature of Nuoc Ngot Lagoon. The EWQI item controls the indicator of water salinity according to the balance equation below:

$$\sigma_{ij} z_{ij} \frac{dC_{ij}}{dt} = \sum_{(i,j)\in\Xi} \left[r_{ij} R_{ij} C_{ij} + E_{ij}\left(C_{ij} - C_{ji}\right)\right] - (1-q_{ij})C_{ij} + F_{ij};$$

where $C_{ij}$ is the concentration of salts in the $\Xi_{ij}$ pixel, $z_{ij}$ is the average depth, $E_{ij}=D_{ij}A_{ij}/L_{ij}$ is the volumetric diffusion coefficient; $L_{ij}$ is the average length of the pixel adjoining boundaries; $A_{ij}$ is the pixel contact area with others pixels; $D_{ij}$ is the turbulent diffusion coefficient; $F_{ij}$ is the source function describing the external inflow of the salts to the $\Xi_{ij}$ pixel (with precipitation, river runoff, from South-China Sea and bottom sediments); $R_{ij}$ is the total water outflow from $\Xi_{ij}$ pixel; $r_{ij}$ is the part of the $R_{ij}$ directed to the border pixels; $\sum_{(i,j)\in\Xi}(r_{ij} + q_{ij}) = 1$; $r_{ij} = 0$ for border-less pixels.

The hydrochemical characteristics of the Nuoc Ngot Lagoon are mainly governed by the water exchange with the South-China Sea across the barrier lagoon strait $\Gamma$ that is characterized by small tidal basin. The masses of water move along the lagoon boundaries $\Gamma$ periodically to the lagoon or from them to the sea. The velocity of this process is evaluated with the following equation (SHP item):

$$V_\Gamma(\varphi,\lambda,t)=V_{\Gamma max}|\cos[\pi(t-t_{max})/\tau]; \ (\varphi,\lambda)\in\Gamma, \ t_{max}-\tau/2\le t\le t_{max}+\tau/2$$

where $\tau$ (≈6.5 hours) is the tide length, $t_{max}$ (11, 17, 24 by local time) is the time when velocity of water flow along $\Gamma$ reaches maximal value $V_{\Gamma max}$ (≈425 m/s).

Changes in the water level in the lagoon depend mainly on the tidal regime and the flow of the river. In addition, the depth of the lagoon is parameterized by the following formula:

$$z_p(\varphi,\lambda,t)=z_0(\varphi,\lambda,t)+\sigma_\Gamma 0.5 V_{\Gamma max} t/\sigma$$

where $\sigma_\Gamma$ ($\approx 45\times10^3$ m$^2$) is the pixel area of the strait $\Gamma$, $z_0$ is the lagoon depth at the end of the tide, and $\sigma$ is the lagoon area.

The external inflow of salts into the lagoon is formed to a greater or lesser extent by the river outflow. Salt sources, such as precipitation and direct outflow of coastal areas, are negligible components that take into account topography (PWL and MCSO items). Hence, the flow $F(\varphi,\lambda,\zeta,\omega,\theta,t)$ of salts of type $\zeta$ in the state $\omega$ with parameter $\theta$ is described by the following function:

$$F(\varphi,\lambda,\zeta,\omega,\theta,t)=\Psi(\varphi,\lambda,\zeta,\omega,\theta,t)Q(\varphi,\lambda,t)\varepsilon(\zeta,\omega,\theta,t)$$

where $\Psi$ is the salt concentration in the soil, $\varepsilon$ is the washout coefficient, $Q$ is the runoff.

## 3. Simulation experiments

Nuoc Ngot Lagoon is linked to the South China Sea from the inlet De Gi and is characterized by the parameters listed in Table 3. The main problem is the optimization of water quality of the lagoon in order to support the maximal productivity of the ecosystem taking into account that lagoon is located in the zone, where man-made activities are strongly developed. The solution to this problem is implemented by monitoring the lagoon, which is based on in-situ measurements at sites No. 1-10 (shown in Fig 4) and in the management of water flow along the $\Gamma$strait.

**Table 3.** Parameters of the Nuoc Ngot Lagoon.

| Parameter | Parameter value | Parameter | Parameter value |
|---|---|---|---|
| Area, km² | 14.7 | Lagoon level variations, cm | -44 - +59 |
| Wind speed, m/s<br>average<br>maximal | <br>1.9<br>39.6 | Temperature of upper layer, °C<br>dry season<br>wet season | <br>26<br>29 |
| Air temperature, °C<br>average<br>minimal<br>maximal | <br>27.0<br>15.8<br>39.9 | Precipitation, mm/year<br>average<br>minimal<br>maximal | <br>1692.9<br>778.0<br>2587.0 |
| Relative air moisture, mb<br>average<br>minimal<br>maximal | <br>27.9<br>20.0<br>32.7 | Distribution of the depths:<br>average depth, m<br>maximal depth, m | <br>1.6<br>9.8 |
| Solar radiation, Kcal/cm² | 144 | Radiation balance, Kcal/cm² | 92.5 |

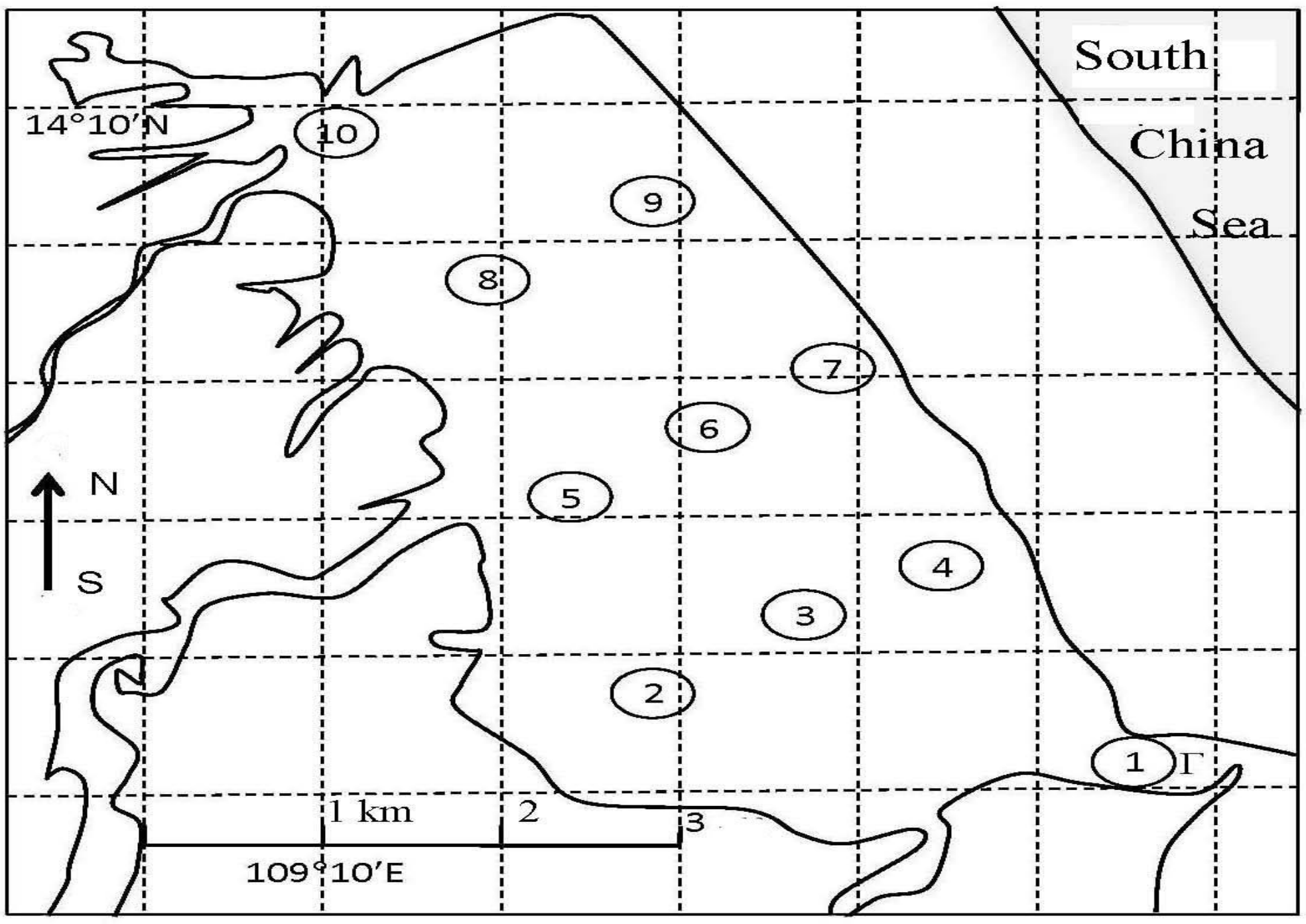

**Fig. 4.** Scheme for the in-situ sites where the water samples are taken.

In-situ measurements are made using the methodology that provides the separation of three water levels: upper, middle and bottom. The chemical characteristics of the water and the analysis of the bottom samples are estimated in the hydrochemical laboratory of the Southern Branch of Vietnam Petroleum Institute (HoChiMinh City). The necessary meteorological information is provided by the Nha Trang weather station.

The model GIMS/NNL allows the spatial reconstruction and forecasting of water quality based on in-situ samples. Tables 4 and 5 illustrate the accuracy of this reconstruction. Simulation experiments show that the correlation between water turbidity and salinity reaches the maximum at 25‰. Bearing this in mind, the variations of water salinity are equal to 7-8% in the Γ strait, 81-85% at river mouths, and 53-55% in other pixels. The commercial importance of the lagoon ecosystem needs the management of water quality, including its salinity. It can be done with the combined use of GIMS/NNL and the control of tidal processes with the barrier management. As shown in Table 4 the optimal water quality level can be achieved when the GIMS/NNL uses data collected only at site No.1 (Γ strait).

**Table 4.** Evaluation of the GIMS/NNL efficiency under the reconstruction of spatial distribution of the Nuoc Ngot Lagoon water quality. Initial data for the simulation experiment is used only from the site No.1. Note: M stands for the modeling results and E for the in-situ measurements.

| Measurement site | Water salinity, °/oo | | Water turbidity, mg/l | | pH | | $PO_4^{-3}$, mg/l | |
|---|---|---|---|---|---|---|---|---|
| | M | E | M | E | M | E | M | E |
| 1 | 30.15 | 33.5 | 12.40 | 10 | 7.04 | 7.82 | 0.033 | 0.03 |
| 2 | 23.14 | 26.0 | 24.78 | 21 | 7.71 | 7.79 | 0.034 | 0.03 |
| 3 | 26.48 | 29.1 | 29.70 | 27 | 7.66 | 7.58 | 0.068 | 0.45 |
| 4 | 27.37 | 32.2 | 26.88 | 24 | 8.01 | 7.78 | 0.023 | 0.02 |
| 5 | 28.53 | 31.7 | 45.59 | 47 | 7.29 | 7.84 | 0.042 | 0.04 |
| 6 | 28.40 | 26.3 | 29.97 | 27 | 7.22 | 7.52 | 0.082 | 0.10 |
| 7 | 27.50 | 25.7 | 23.94 | 21 | 7.45 | 7.30 | 0.067 | 0.35 |
| 8 | 26.16 | 25.4 | 34.58 | 38 | 7.21 | 7.75 | 0.023 | 0.02 |
| 9 | 26.54 | 30.5 | 48.45 | 51 | 7.09 | 7.71 | 0.023 | 0.02 |
| 10 | 20.95 | 26.5 | 60.48 | 63 | 7.63 | 7.00 | 0.039 | 0.05 |
| Average error, % | 10.7 | | 11.0 | | 6.0 | | 15.4 | |

**Table 5.** Assessment of the GIMS/NNL precision when the Nuoc Ngot Lagoon water salinity is predicted. Initial data at the time $t_0$ is used from the site No.1.

| Site | In-situ measurement of the water salinity (°/oo) at the time $t_0$ | Prognosis and error (%) | | | | |
|---|---|---|---|---|---|---|
| | | $t_0$+ 7 days | $t_0$+ 14 days | $t_0$+ 21 days | $t_0$+ 1 month | $t_0$+ 1.5 month |
| 1 (strait Г) | 30.2 | 32.0(5) | 32.8(9) | 26.4(10) | 29.3(12) | 34.4(14) |
| 2 | 33.6 | 32.8(2) | 34.4(10) | 27.6(15) | 26.4(19) | 25.7(23) |
| 3 | 33.6 | 32.4(5) | 33.7(12) | 37.5(17) | 25.9(19) | 26.2(22) |
| 4 | 33.7 | 33.4(4) | 28.3(12) | 37.8(17) | 26.6(20) | 25.6(24) |
| 5 | 33.8 | 33.4(4) | 28.8(11) | 27.6(17) | 25.2(20) | 26.5(22) |
| 6 | 33.8 | 31.0(8) | 30.9(12) | 27.8(17) | 27.5(20) | 25.3(25) |
| 7 | 30.7 | 31.4(3) | 34.9(13) | 27.9(17) | 30.2(9) | 34.5(19) |
| 8 | 32.4 | 34.7(7) | 30.9(9) | 34.9(16) | 26.6(19) | 25.2(22) |
| 9 | 32.5 | 33.0(6) | 29.4(10) | 28.6(18) | 28.3(20) | 25.1(23) |
| 10 | 30.4 | 31.9(9) | 33.2(12) | 34.5(17) | 33.2(10) | 27.8(24) |

As can be seen from Tables 4 and 5, the spatial distribution of water salinity is reconstructed in two weeks with a precision of no more than 84%. The biggest error occurs at sites 3 and 7 for $PO_4^{3-}$. Cause of this is the observed anomalous emission of chemicals from the sediments at sites 3 and 7, where the correction of model coefficients in the items MTM, SHP and SEC is required in the first position.

However, as can be seen from Figure 5, a precision of 85-90% can be achieved with in-situ measurements at site 1 with a frequency of one month and using GIMS/NNL for the calculations of water quality characteristics during this month. Undoubtedly, this result depends on the reliability of meteorological forecast. Figure 6 shows the results of the simulation experiments, when different meteorological conditions are considered. It is noted that the precision of the forecast stabilizes during the dry season while the wet season is characterized by the lower reliability of the modeling results. The average forecasting error during the dry season was 9.7% less, with the exception of 2010, when it exceeded 10.6%. Wet seasons are characterized by higher forecast errors due to all errors in meteorological data and estimates of coastal and river outflows. Nevertheless, predicting the water salinity and pH is characterized by errors of the range of 10.4%, regardless of the season.

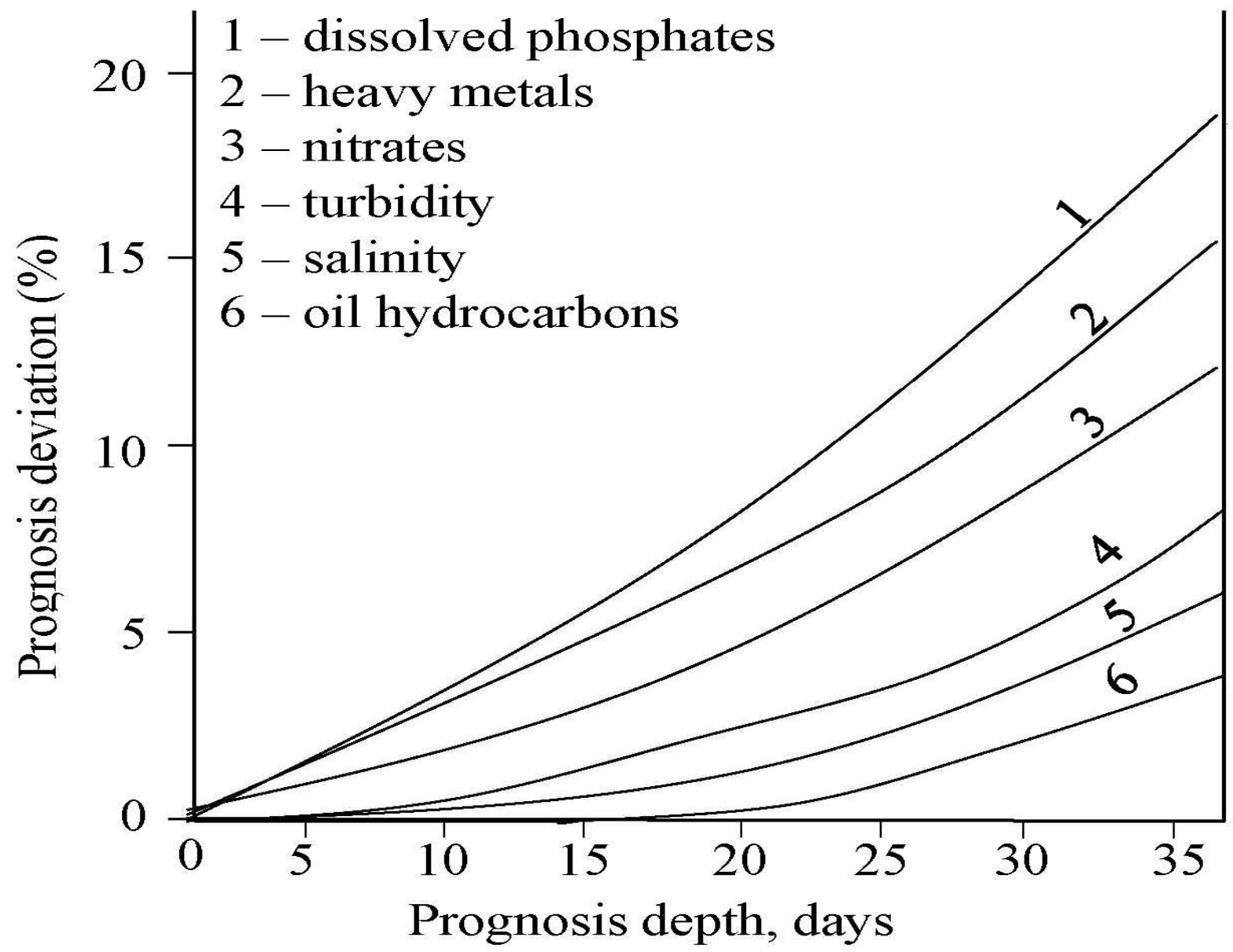

**Fig. 5.** Dependence of prediction deviation on depth of forecasting time and on water quality characteristics.

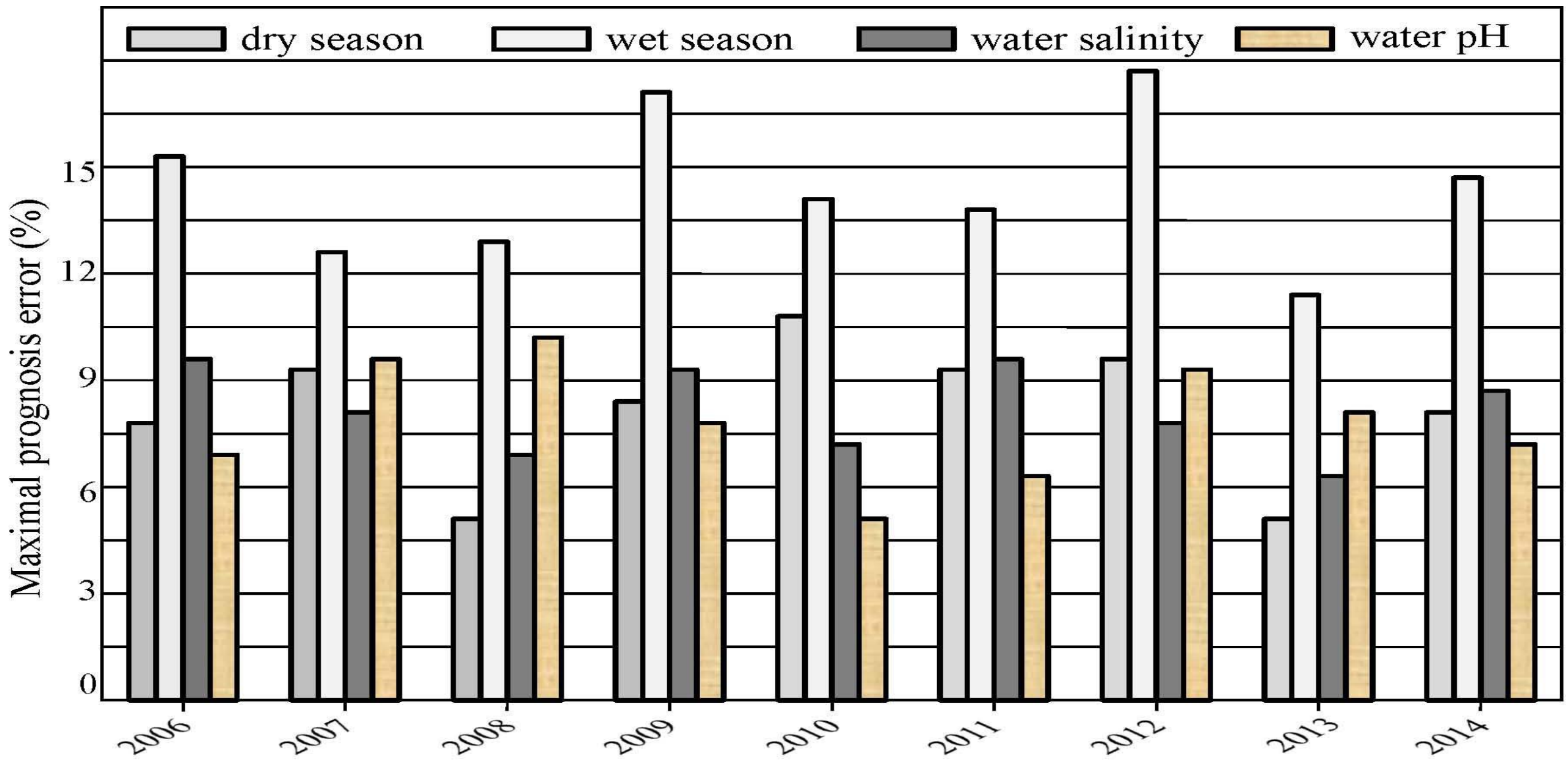

**Fig. 6.** Temporal evolution of the maximum error of the water quality forecast at the Nuoc Nuno lagoon through the GIMS / NNL occurring each month during the year in which the original data is measured at location 1 and the errors are evaluated at positions 2-10. .

The optimal ecological conditions in the Nuoc Ngot Lagoon zone are achieved through tidal management, when other factors such as contaminant flows in rivers, precipitation and coastal runoff are controlled. Table 6 shows the dependence of the water salinity due to tidal processes. The average ecologically acceptable salinity of water ranges between 22.5°/oo and 24.5°/oo. Reducing the inflow of sea water into the lagoon leads to a decrease in the salinity of the water due to the outflow of the river.

As shown in Table 6 the optimal regime of water inflow management is consistent with the water inflow in the lagoon in the range of 300-500 $m^3/s$. The duration of the experiment is one, two and three weeks. When the strait Γ closed, the salinity of lagoon water is reduced during one week from 23.5°/oo to 20.01°/oo during the dry season and continues to decrease up to 16.34°/oo the following three weeks. When $V_\Gamma$>500 $m^3/s$ water salinity increases rapidly during both wet and dry seasons. The latter states that river inflows do not balance the salinity of water without supporting it with an average value of 23.5°/oo (Bui Quoc Nghia, 2002).

**Table 6.** A dependence of the lagoon water salinity on the volume of the sea water inflow into the lagoon.

| Velocity of the sea water inflow, $V_Γ$ ($m^3/s$) | Dry season | | | Wet season | | |
|---|---|---|---|---|---|---|
| | 1 week | 2 weeks | 3 weeks | 1 week | 2 weeks | 3 weeks |
| 0 | 20.01 | 18.22 | 16.34 | 19.36 | 16.86 | 14.77 |
| 50 | 21.41 | 22.03 | 22.54 | 20.13 | 21.16 | 22.22 |
| 100 | 22.11 | 23.05 | 23.49 | 21.29 | 21.54 | 22.62 |
| 200 | 22.74 | 22.95 | 23.51 | 21.71 | 22.23 | 23.47 |
| 300 | 23.53 | 23.64 | 23.65 | 22.97 | 23.48 | 23.52 |
| 400 | 23.55 | 23.65 | 23.67 | 23.31 | 23.42 | 23.53 |
| 500 | 24.92 | 24.97 | 25.24 | 23.78 | 23.89 | 24.56 |
| 600 | 27.58 | 28.55 | 30.94 | 27.12 | 28.34 | 30.49 |

## 4. Conclusions and perspectives

From the analysis and discussion mentioned above, it appears that the implementation of water quality control in a regional or local hydrochemical field (e.g. a lagoon) requires the optimization task to be involved in the economic and instrumental supply of this function. In this respect, GIMS technology allows this task to be solved using a combination of mathematical and instrumental tools. The Nuoc Ngot Lagoon is a typical mesoscale hydrochemical coastal area, where hydrological processes are determined by the river outflow and the water exchange with the South China Sea. Simulation experiments have shown that in this case it is possible to optimize water quality control through regular in-situ measurements and simulations. The most important component of this process is the up-to-date information on anthropogenic interference in the lagoon area.

Thus, the GIMS/NNL allows the optimization of lagoon monitoring through the reduction of in-situ measurements over time and the number of sites. Reliable monitoring results are achieved when in-situ measurements are made at site 1 (Γstrait) with a three-week period in the dry season and two weeks in a wet period. The monitoring process can be automated using an adaptive spectroellipsometric sensor at the fixed position, which allows the evaluation of water quality characteristics in real time and the management of the tidal regime (Krapivin and Mkrtchyan 2016; Mkrtchyan and Krapivin 2016). The implementation of this task is pending and will be explored by the same authors in the near future.

## Acknowledgments

This study partly was supported by the Russian Foundation for Basic Research, Project No. 16-01-000213-a.

## References

[1] Chattopadhyay, G., Chakraborthy, P. & Chattopadhyay, S. (2012). Mann-Kendall trend analysis of tropospheric ozone and its modeling using ARIMA, *Theoretical and Applied Climatology, 110*, 321–328, doi:10.1007/s00704-012-0617-y.

[2] U.S. Environmental Protection Agency, 2007, EPA relying on existing clean air act regulations to reduce atmospheric deposition to the Chesapeake Bay and its watershed: Report Number 2007-P-00009, 13 p. plus appendices.

[3] Xue, Y., He, X.W., Xu, H., Guang, J., Guo, J.P. & Mei, L.L. (2014). China Collection 2.0: The aerosol optical depth dataset from the synergetic retrieval of aerosol properties algorithm, Atmospheric Environment, 95, 45–58, doi:10.1016/j.atmosenv.2014.06.019.

[4] Varotsos, C.A. & Cracknell, A. P. (1993). Ozone depletion over Greece as deduced from Nimbus-7 TOMS measurements. International Journal of Remote Sensing, 14(11), 2053-2059.

[5] Varotsos, C.A. & Cracknell, A.P. (1994). Three years of total ozone measurements over Athens obtained using the remote sensing technique of a Dobson spectrophotometer. International Journal of Remote Sensing, 15(7), 1519-1524.

[6] Varotsos, C.A. & Cracknell, A.P. (2004). New features observed in the 11-year solar cycle. International Journal of Remote Sensing, 25(11), 2141-2157.

[7] Kondratyev, K.Y., Varotsos, C.A., & Cracknell, A.P. (1994). Total ozone amount trend at St Petersburg as deduced from Nimbus-7 TOMS observations.International Journal of Remote Sensing, 15(13), 2669-2677.

[8] Reid, S.J., Rex, M., Von Der Gathen, P., Fløisand, I., Stordal, F., Carver, G.D., ... & Braathen, G. (1998). A Study of Ozone Laminae Using Diabatic Trajectories, Contour Advection and Photochemical Trajectory Model Simulations. Journal of Atmospheric Chemistry, 30(1), 187-207.

[9] Feretis, E., Theodorakopoulos, P., Varotsos, C., Efstathiou, M., Tzanis, C., Xirou, T., ... & Aggelou, M. (2002). On the plausible association between environmental conditions and human eye damage. Environmental Science and Pollution Research, 9(3), 163-165.

[10] Varotsos, C., Kondratyev, K. Y., & Katsikis, S. (1995). On the relationship between total ozone and solar ultraviolet radiation at St. Petersburg, Russia. Geophysical Research Letters, 22(24), 3481-3484.

[11] Cracknell, A.P., & Varotsos, C.A. (1994) Ozone depletion over Scotland as derived from Nimbus-7 TOMS measurements. International Journal of Remote Sensing, 15(13), 2659-2668.

[12] Cracknell, A.P., & Varotsos, C.A. (1995) The present status of the total ozone depletion over Greece and Scotland: a comparison between Mediterranean and more northerly latitudes. International Journal of Remote Sensing, 16(10), 1751-1763.

[13] Efstathiou, M., Varotsos, C., & Kondratyev, K.Y. (1998). An estimation of the surface solar ultraviolet irradiance during an extreme total ozone minimum. Meteorology and Atmospheric Physics, 68(3), 171-176.

[14] Cracknell, A.P., & Varotsos, C.A. (2007). Editorial and cover: Fifty years after the first artificial satellite: from sputnik 1 to envisat.

[15] Varotsos, C., Kalabokas, P., & Chronopoulos, G. (1994). Association of the laminated vertical ozone structure with the lower-stratospheric circulation. Journal of Applied Meteorology, 33(4), 473-476, 1994.

[16] Gernandt, H., Goersdorf, U., Claude, H., & Varotsos, C.A. (1995). Possible impact of polar stratospheric processes on mid-latitude vertical ozone distributions. International Journal of Remote Sensing, 16(10), 1839-1850.

[17] Varotsos, C.A. & Tzanis, C. (2012). A new tool for the study of the ozone hole dynamics over Antarctica, Atmos. Environ., 47, 428–434.

[18] Varotsos, C. (2002). The southern hemisphere ozone hole split in 2002. Environmental Science and Pollution Research, 9(6), 375-376.

[19] Efstathiou, M. N., Varotsos, C. A., Singh, R. P., Cracknell, A. P., & Tzanis, C. (2003). On the longitude dependence of total ozone trends over middle-latitudes. International Journal of Remote Sensing, 24(6), 1361-1367.

[20] Reid, S.J., Vaughan, G., Mitchell, N.J., Prichard, I.T., Smit, H.J., Jorgensen, T.S., ... & De Backer, H. (1994). Distribution of ozone laminae during EASOE and the possible influence of inertia-gravity waves. Geophysical Research Letters, 21(13), 1479-1482.

[21] Varotsos, C.A., Kondratyev, K.Y. & Cracknell, A.P. (2000). New evidence for ozone depletion over Athens, Greece. International Journal of Remote Sensing, 21(15), 2951-2955.

[22] Varotsos, C. & Cartalis, C. (1991). Re-evaluation of surface ozone over Athens, Greece, for the period 1901–1940. Atmospheric Research, 26(4), 303-310.

[23] Efstathiou, M.N., Tzanis, C., Cracknell, A.P. & Varotsos, C.A (2011). New features of land and sea surface temperature anomalies. International journal of Remote Sensing, 32(11), 3231-3238.

[24] Varotsos, C. (2005). Modern computational techniques for environmental data; application to the global ozone layer. Computational Science–ICCS 2005, 43-68.

[25] Varotsos, C., Efstathiou, M. & Tzanis, C. (2009). Scaling behaviour of the global tropopause. Atmospheric Chemistry and Physics, 9(2), 677-683.

[26] Efstathiou, M.N. & Varotsos, C.A. (2010). On the altitude dependence of the temperature scaling behaviour at the global troposphere. International Journal of Remote Sensing 31, 343-349.

[27] Varotsos, C.A., Tzanis, C.G., & Sarlis, N.V. (2016). On the progress of the 2015–2016 El Niño event. Atmospheric Chemistry and Physics, 16(4), 2007-2011.

[28] Kondratyev, K. & Varotsos, C. (1995). Atmospheric greenhouse effect in the context of global climate change. Il Nuovo Cimento C, 18(2), 123-151.

[29] Cracknell, A.P. & Varotsos, C.A. (2011). New aspects of global climate-dynamics research and remote sensing. International Journal of Remote Sensing 32(3), 579-600.

[30] Krapivin, V.F. & Varotsos, C.A. (2016). Modelling the CO2 atmosphere-ocean flux in the upwelling zones using radiative transfer tools. Journal of Atmospheric Solar-Terrestrial. Physics, 150-151, 47-54.

[31] Mai Trong, Nhuan, Tran Hong, Ha, Nguyen Chi, Thanh,…, Hoang Van, Thang Pham Dinh & Viet Hong (2005). Overview of wetlands status in Vietnam following 15 years of RAMSAR Convention implementation. Vietnam Environment Protection Agency, Hanoi, 73pp.

[32] McElwee, P. (2010). The social dimensions of adaptation to climate change in Vietnam. The World Bank, New York, 140 pp.

[33] Piazza, R., Giuliani, S., Bellucci, L.G., Mugnai, C., Nguyen, Huu Cu, Dang, Hoai Nhon, Vecchiato, M., Romano, S. & Frignani, M. (2010). PCDD/Fs in sediments of Central Vietnam coastal lagoons: In search of TCDD, Marine Pollution Bulletin, 60, 2303–2310.

[34] Krapivin, V.F., Varotsos, C.A. & Soldatov, V.Yu. (2015). New Ecoinformatics Tools in Environmental Science: Applications and Decision-making. Springer, London, U.K., 903 pp.

[35] Rodriguez-Rodriguez, M. & Moreno-Ostos, E. (2006). Heat budget, energy storage and hydrological regime in a coastal lagoon, Limnologia, 36, 217-227.

[36] Ferrarin, G., Zaggia, L., Paschini, E. & Guerzoni, S. (2013). Hydrological regime and renewal capacity of the micro-tidal Lesina Lagoon, Estuaries and Coasts, 37 (1), 79-93.

[37] Jakimavičius, D. & Kovalenkovienė, M, (2010). Long–term water balance of the Curonian Lagoon in the context of anthropogenic factors and climate change, Baltica (Vilnius), 23 (1): 33-46.

[38] Tran, Thanh Tung (2011). Morphodynamics of seasonally closed coastal inlets at the coast of Vietnam. Ipskamp Drukkers, The Netherlands, 191 pp.

[39] Tran, Van Son (2016). Towards successful implementation of Vietnamese national government climate change policy at the provincial and local farmer level, PhD Theses, Southern Cross University, Lismore, 290 pp.

[40] Bui Quoc Nghia (2002). Simulation system for the hydrophysical experiment in the anisotropic environment. PhD theses. Moscow Oceanology Institute of Russian Academy of Sciences, Moscow, 151 pp [in Russian].

[41] Cribb, J. (2017). Surviving the 21 st century. Surviving the 21st Century: Humanity's Ten Great Challenges and How We Can Overcome Them. Springer, Switzerland, 255 pp., doi 10.1007/978-3-319-41270-2.

[42] Zhengtao, L. (2015). Water quality criteria green book of China. Springer, The Netherlands, 161 pp.

[43] Romano, S., Mugnai, C., Giulian,i S., Turetta, C., Nguyen, Huu Cu, Bellucci, L.G., Dang, Hoai Nhon, Capodaglio, G., Frignani M. (2012). Metals in Sediment Cores from Nine Coastal Lagoons in Central Vietnam, American Journal of Environmental Sciences 8, (2), 130-142.

[44] Romano, S., Mugnai, C., Giuliani, S., Nguyen, Huu Cu, Bellucci, L.G., Turetta, C., Capodaglio, G., Dang, Hoai Nhon, Albertazzi, S. & Frignani, M. (2013). Extreme events and environmental changes: Tracing sedimentary processes in Central Vietnam coastal lagoons, Chemistry and Ecology, 29 (2), 166-180.

[45] Wells, N.C. (2011). The atmosphere and ocean: A physical introduction. Wiley, New York, 424 pp.

[46] Krapivin, V. F., & Varotsos, C. A. (2008). Biogeochemical cycles in globalization and sustainable development. Springer/Praxis, Chichester.